\documentclass[referee,pdflatex,sn-mathphys,Numbered]{sn-jnl}

\usepackage{graphicx}%
\usepackage{multirow}%
\usepackage{amsmath,amssymb,amsfonts}%
\usepackage{amsthm}%
\usepackage{mathrsfs}%
\usepackage[title]{appendix}%
\usepackage{xcolor}%
\usepackage{textcomp}%
\usepackage{manyfoot}%
\usepackage{booktabs}%
\usepackage{listings}%

\begin{document}

\title[Is $d^*(2380)$ a compact hexaquark state?]{Is $d^*(2380)$ a compact hexaquark state?}

\author[1]{\fnm{Manying} \sur{Pan}}\email{211001005@njnu.edu.cn}
\author[2]{\fnm{Xinmei} \sur{Zhu}}\email{zxm\_yz@126.com}
\author*[1]{\fnm{Jialun} \sur{Ping}}\email{jlping@njnu.edu.cn}

\affil[1]{\orgdiv{Department of Physics and Jiangsu Key Laboratory for Numerical Simulation of Large Scale Complex Systems},
\orgname{Nanjing Normal University}, \orgaddress{\city{Nanjing}, \postcode{210023}, \country{P. R. China}}}
\affil[2]{\orgdiv{Department of Physics}, \orgname{Yangzhou University}, \orgaddress{\city{Yangzhou}, \postcode{225009}, \country{P.R. China}}}

\abstract{The most fascinating dibaryon in the non-strange quark sector is $d^*(2380)$,
which was reported by WASA-at-COSY Collaboration and confirmed by A2@MAMI Collaboration.
The reported mass and decay width are $M\approx2.37$ GeV,
$\Gamma\approx70$ MeV and the quantum numbers are $IJ^P=03^+$.
The structure of $d^*(2380)$ is still in controversy.
In the present calculation,
the powerful method in few-body system,
Gaussian Expansion Method (GEM) is employed to explore the structure of $d^*(2380)$
in the framework of constituent quark models
without assuming the presupposed structure.
The results show that the radius of $d^*(2380)$ is around $0.8$ fm,
it is a very compact object. Because of the compact structure,
the color singlet-singlet component has a large overlap with the color octet-octet one,
two colorless, large overlapped $\Delta$s dominate the state is possible.}

\keywords{dibaryon, quark model, Gaussian expansion method, hidden-color channel}

\maketitle

\section{Introduction}
In addition to the popular XYZ particles and the hidden charm pentaquarks,
the dibaryon states are also important exotic hadron states and worth profound study.
Generally speaking,
any object with a baryon number $B=2$ can be called a dibaryon.
Since the baryon number of each quark is 1/3,
the dibaryon is composed of six valence quarks.

The proposal of looking for dibaryons was in the same year as the publication of quark model by Gell-Mann~\cite{Gell-Mann:1964ewy}.
In 1964,
based on $SU(6)$ symmetry of strong interaction,
Dyson and Xuong predicted the possible existence of dibaryon states and obtained the mass of these particles by a mass formula~\cite{Dyson:1964xwa},
the predicted mass of $D_{03}$ is surprisingly close to that of $d^*(2380)$ later
found~\cite{PRL102,WASA-at-COSY:2011bjg,WASA-at-COSY:2014qkg,WASA-at-COSY:2014lmt,WASA-at-COSY:2014dmv,Adlarson:2016bxt}.

Deuteron is a state with $B=2$,
which was discovered by Urey, Brickwedde and Murphy in 1932~\cite{Urey:1932gik}.
It is a loosely bound state of proton and neutron with quantum numbers $IJ=01$.
At the quark level,
the content of deuteron is $uuuddd$,
these six quarks could also make up $\Delta\Delta$,
so whether deuteron contains non-nucleon component
and its internal structure are meaningful subjects~\cite{Julia-Diaz:2002zuf,Glozman:1994xe,Weinberg:1965zz}.
Deuteron is currently the only confirmed stable dibaryon system,
and PDG list its mass as a physical constant~\cite{Workman:2022ynf}.
Due to the large separation between proton and neutron in deuteron~\cite{DeVries:1987atn},
it can be safely regarded as a molecular state.
Of course,
the dibaryon state may also be a more exotic compact six quark structure,
that is,
the state cannot be represented by two well separated color singlet quark clusters.
It is more interesting because it is a new form of matter.
$d^*$ with quantum numbers $IJ^P=03^+$ is expected to be a compact object in quark model calculations~\cite{Ping:2000dx,Huang:2019lzt}.
Although many possible states are predicted in theory,
experimental dibaryon search experienced a long and eventful history,
there are many twists and turns during the searches of dibaryons,
a comprehensive review of dibaryons can be found in the references~\cite{Clement:2016vnl,Clement:2020mab},
the experimental status of $d^*$ can be seen in~\cite{Skorodko:2017ttk}.

After the initial prediction of $d^*$ in 1964,
the further study of dibaryon state related to $d^*$ was traced back to 1977.
Inspired by the anomalous results of proton polarization in the $\gamma d \rightarrow pn $ reaction~\cite{Kamae:1976at2},
Kamae and Fujita investigated the possible existence of deep bound dibaryon state,
in which they calculated the $\Delta\Delta$ state with quantum numbers $IJ=03$ and $IJ=30$ using the non relativistic one boson exchange model,
and obtained a binding energy of about 100 MeV~\cite{Kamae:1976at}.
In fact, the results of these early researches are coincided with that of the $d^*$ later found in the
experiments~\cite{PRL102,WASA-at-COSY:2011bjg,WASA-at-COSY:2014qkg,WASA-at-COSY:2014lmt,WASA-at-COSY:2014dmv,Adlarson:2016bxt}.

In 1989, Goldman {\em et al} proposed ``an inevitable nonstrange dibaryon"~\cite{Goldman:1989zj},
which was named $d^*$.
The following realistic calculations in quark delocalization and color screening model(QDCSM) confirmed the prediction of
$d^*$~\cite{Wang:1995bg,Ping:2000cb,Ping:2000dx}.
In the framework of chiral quark model,
the results have shown that there are attractions between two $\Delta$'s,
the dynamical calculation with the help of the resonating group method (RGM) obtained small binding energy, $22.2-64.8 $ MeV for $d^*$ and a compact structure,
the root-mean-square radius (RMS) is about $0.84-1.01$ fm~\cite{NPA683}.
To guide the experimental searching,
a nucleon-nucleon ($NN$) scattering phase shifts calculation including $d^*$ was performed,
the phase shifts of $D$-wave $NN$ scattering show a clear resonance structure,
with the mass $2273-2404 $ MeV and width $33-149$ MeV ~\cite{Ping:2008tp},
the decay width is associated with the ABC effect,
which is named after its first discoverer Abashian, Booth,and Crowe~\cite{Abashian:1960zz}.
The experimental breakthrough occurs in 2009,
CELSIUS/WASA-at-COSY Collaboration reported their results on double pionic fusion reaction $pn\rightarrow d\pi^0\pi^0$,
a resonance with mass and width 2.36 GeV and 80 MeV is needed to describe the
experimental data~\cite{PRL102},
the subsequent series of experiments confirmed the resonance and fixed quantum numbers~\cite{WASA-at-COSY:2011bjg,WASA-at-COSY:2014qkg,WASA-at-COSY:2014lmt,WASA-at-COSY:2014dmv}.
The updated results are the resonance mass is around 2.37 MeV,
the decay width is about 70 MeV and quantum numbers are $IJ^P=03^+$.
It is a dibaryon $d^*$,a spin excitation of deuteron.

Especially,
the re-analysis of $NN$ scattering amplitude in $^3D_3$-$^3G_3$ partial waves
by incorporating new data suggest a pole which corresponding to $d^*$~\cite{WASA-at-COSY:2014dmv}.
The theoretical study of a $d^*$ resonance in the coupled
$^3D_3$-$^3G_3$ partial waves of $NN$ scattering reproduced the experimental data~\cite{Huang:2014aca}.
A dynamical calculation of the $\Delta\Delta$ dibaryon candidates under the quark delocalization color screening model and the chiral quark model obtained similar results,
their results show that the attractions between two $\Delta$s is strong enough to
bind two $\Delta$s together,
introduction of the hidden-color channels in ChQM will lowered resonance masses by $10-20$ MeV~\cite{Huang:2013nba},
which is consistent with the results of this paper.
The recent polarization experiment of A2 Collaboration at MAMI also find signatures of the $d^*(2380)$ hexaquark in d($\gamma$,$p\vec{n}$)~\cite{A2:2019arr}.
For the signals invoking the existence of dibaryon in WASA-at-COSY experiments, there are also other explanations without dibaryon.
Ikeno \emph{et al}. proposed triangle singularity to explain the experimental data~\cite{Oset1,Oset2}.

After the experiment discovery,
more researches are devoted to the structure and the narrow decay width of $d^*$.
To understand the narrow decay width of $d^*$,
the assumption that the dominant component of $d^*$ is hidden color channel was proposed~\cite{Bashkanov}.
The assumption comes from the transformation between the physical bases (denoted by two $q^3$ state) and the symmetry bases (denoted by the orbital symmetry and isospin-spin symmetry)~\cite{Harvey,Wang:1995kp}.
\begin{table}
\caption{\label{tab:table1}The transformation coefficients between physical bases and symmetry bases. $[\nu]$ and $[\mu]$ denote the symmetry of orbital
and spin-flavor for six-quark systems.}
\begin{tabular}{ccc}
\toprule
    & ~~~~$[\nu][\mu]=[6][33]$~~~~  & ~~~~$[\nu][\mu]=[42][33]$~~~~ \\
 $\Delta\Delta$ & $-\sqrt{1/5}$  &   $-\sqrt{4/5}$ \\
 $CC$           & $-\sqrt{4/5}$  &   $\sqrt{1/5}$ \\
\botrule
\end{tabular}
\end{table}
From the table~\ref{tab:table1},
one can see that if the orbital symmetry of $d^*$ is $[6]$, which is the case with only one orbital single particle state available,
for example, six quarks are put into one bag, in the symmetry bases, then according to the coupling among orbital, color, flavor and spin between
all six quarks occupy the same orbital state,
then in the physical bases,
the hidden color channel (CC) is the dominant component (80\%).
Really in the resonating-group-method (RGM) approach,
by including the hidden-color channel (CC),
the calculation of $d^*$ in chiral quark model gave that $d^*$ has a mass of about $2.38-2.42$ GeV and
a root-mean-square radius (RMS) of about $0.76-0.88$ fm,
and the fraction of CC component in the $d^*$ is found to be about
66$\%$-68$\%$~\cite{Huang:2019lzt,Huang:2016zox,Huang:2014kja}.
However,
there is a mis-understanding of the above transformation table~\ref{tab:table1}.
If only one orbital single particle state available, for example, six quarks are put into one bag and all in the same orbital state, then the orbital symmetry
of six-quark state is limited to $[6]$, the orbital symmetry $[42]$ will disappear. Then according to the coupling among orbital $[6]$, color $[222]$,
flavor $[33]$ and spin $[6]$, $[6]\times [222]\times [33]\times [6]=[1^6]$, only one basis in the symmetry bases scheme is available.
Then the corresponding available physical basis must be one, too,
the color-singlet channel $\Delta\Delta$ is the same as the hidden-color channel $CC$.
In the RGM approach,
the overlap between $\Delta\Delta$ and $CC$ is about 1 when the separation between two clusters are small,
for example $\langle \Delta\Delta|CC\rangle =0.98$
with separation $s=0.5$ fm~\cite{dstar2015beijing}.
Very recently,
Huang performed a revised quark model investigation of $d^*(2380)$~\cite{Huang:2022qgx},
and pointed out that there are some inadequacies in their previous quark model calculations,
it would be imprecise to set size parameter to be same for
all the considered baryons,
and accordingly the coupling strengths of one-gluon-exchange (OGE) potential were not well determined.
In the updated the chiral quark model calculation,
the author found the effects of hidden-color channel are much less important,
which is different from their previous work~\cite{Huang:2019lzt,Huang:2016zox,Huang:2014kja}.

As for the structure of $d^*$, the most quark model calculation show that it is compact object~\cite{NPA683,Ping:2008tp,Lu:2017uey}.
However, in the three-body Faddeev equation approach of $\pi N\Delta$, the extended object is invoked to explain $d^*$~\cite{Gal:2013dca,Gal:2014zia}.
In lattice QCD approach, the similar results with that of quark model calculations are obtained, the short-range strong attraction between two
$\Delta$s leads to the quasi-bound states with compact structure~\cite{Gongyo:2020pyy}.

In quark model calculations, RGM is often employed. It is an approximation method for few-body system, in which the system is separated into
two sub-clusters and the structures of the sub-clusters are frozen in the dynamical calculation. In this way, the multi-body problem was simplified
into two-body one. It is expected to be a good approximation in nucleon-nucleon scattering study. It maybe not suitable for studying the structure
and the percentage of hidden-color channel in $d^*$. In the present work, the powerful method in few-body problem,
gaussian expansion method (GEM)~\cite{GEM,Hiyama:2018ivm} is invoked to determine the contribution of hidden-color channels and root-mean-square radius of $d^*$.

This paper is structured as follows:
Sec.II briefly introduced the quark models,
the construction of hexaquark wave functions and GEM.
The calculated results and discussions are presented in Sec.III.
The summary of our investigation is given in the last section.

\section{Models and wave functions}
To check the model dependence of the calculation, two quark models are used, one is the na\'{i}ve quark model, another is the chiral
quark model.
The calculations are limited to the ground states, so only the central parts of Hamiltonian are given below.
\subsection{Na\'ive Quark Model}
In the na\'{i}ve quark model, the interaction between quarks occurs by exchanging a gluon.Hamiltonian includes the static mass of all constituent quarks,kinetic energy term, color confinement potential and one gluon exchange potential,which can be written as:
\begin{eqnarray}
H & = & \sum_{i=1}^{6}\left( m_i+\frac{p^2_i}{2m_i}\right)-T_{CM}
      + \sum_{j>i=1}^{6} V_{ij}, \\
V_{ij} & = & V^{C}_{ij}+V^{G}_{ij}, \nonumber \\
V^{C}_{ij}
& = &  -a_c  \mbox{\boldmath $\lambda$}_i^c\cdot\mbox{\boldmath $\lambda$}_j^c(r^2_{ij}+V_0),  \\
V^{G}_{ij}
& = & \frac{\alpha_s}{4}\mbox{\boldmath $\lambda$}_i^c\cdot\mbox{\boldmath $\lambda$}_j^c
       \left[ \frac{1}{r_{ij}}-\frac{\mbox{\boldmath $\sigma$}_i\cdot\mbox{\boldmath $\sigma$}_j}{6m_im_j}
              \frac{e^{-r_{ij}/r_0(\mu)}}{r_{ij}r^2_0(\mu)}\right] , \\
       & &        ~~~r_0(\mu)=\hat{r}_0/\mu. \nonumber
\end{eqnarray}
Where $m_i$ is the constituent mass of quark, $p_i$ is momentum of quark, $T_{CM}$ is center of mass kinetic energy, $V^{C}_{ij}$ means color confinement potential,
$V^{G}_{ij}$ stands for one-gluon exchange potential (OGE), {\boldmath $\lambda$} and {\boldmath $\sigma$} are $SU(3)$  Gell-Mann color and $SU(2)$ Pauli spin matrices respectively,
$\mu$ represents the reduced mass between two interacting quarks.

\subsection{Chiral Quark Model}
Chiral quark model was setup based on the dynamic breaking of chiral symmetry~\cite{Valcarce:1995dm}.
Due to chiral symmetry spontaneous breaking,
Goldstone boson exchange potentials appear between light quarks,
pseudoscalar $(\pi)$ and scalar $(\sigma)$ meson exchange terms are invoked,
in addition to the color confinement
and one-gluon-exchange potentials.
The Hamiltonian in chiral quark model is written as~\cite{Vijande:2004he}:
\begin{eqnarray}
H &=& \sum_{i=1}^6 \left(m_i+\frac{p_i^2}{2m_i}\right) -T_{CM} \nonumber \\
& & +\sum_{i<j} \left[ V^{G}_{ij}+V^{\pi}_{ij}+V^{\sigma}_{ij}+V^{C}_{ij} \right],  \\
V^{\pi}_{ij}&=& \frac{1}{3}\alpha_{ch}
\frac{\Lambda^2}{\Lambda^2-m_{\pi}^2}m_\pi \left[ Y(m_\pi
r_{ij})- \frac{\Lambda^3}{m_{\pi}^3}Y(\Lambda r_{ij}) \right] \nonumber \\
& & {\mbox{\boldmath $\sigma$}}_i \cdot{\mbox{\boldmath $\sigma$}}_j {\boldsymbol \tau}_i \cdot {\boldsymbol \tau}_j,  \nonumber \\
V^{\sigma}_{ij}&=& -\alpha_{ch} \frac{4m_u^2}{m_\pi^2}
\frac{\Lambda^2}{\Lambda^2-m_{\sigma}^2}m_\sigma \nonumber \\
& & \left[ Y(m_\sigma
r_{ij})-\frac{\Lambda}{m_\sigma}Y(\Lambda r_{ij})
\right],~~~\alpha_{ch}= \frac{g^2_{ch}}{4\pi}\frac{m^2_{\pi}}{4m^2_u}  \nonumber
\end{eqnarray}
Where $V^{\pi}_{ij}$ and $V^{\sigma}_{ij}$ represent one $\pi$ and one $\sigma$ exchange potentials. $Y(x)$ is standard Yukawa functions,
$Y(x)=\frac{e^{-x}}{x}$, $\alpha_{ch} $ is the chiral coupling constant between quark and Goldstone bosons,
which is determined as usual from the $\pi$-nucleon coupling constant~\cite{Fernandez:1993hx,Obukhovsky:1990tx}.
Other symbols have their usual meanings.

With this model,
not only the properties of baryons and mesons can be well described,
but also the existing experimental data of deuteron and $NN$ scattering can be well described,
it has been used to study few-baryon systems~\cite{Valcarce:2005em}.

\subsection{Wave functions}
The quark has four degrees of freedom: orbital, spin, color, and flavor.
We construct the wave functions for each degree of freedom as follows. For each degree of freedom,
the six-quark system is separated into two sub-clusters,
$a$ (quarks 1, 2 and 3) and $b$ (quarks 4, 5 and 6).
First we construct the wavefunctions for each sub-cluster,
then couple two wavefunctions of two sub-clusters to get the total wavefunctions for a six-quark system.

\noindent{1. orbital wave functions}

\begin{figure}[htbp]
\begin{center}
\includegraphics[width=3.5in]{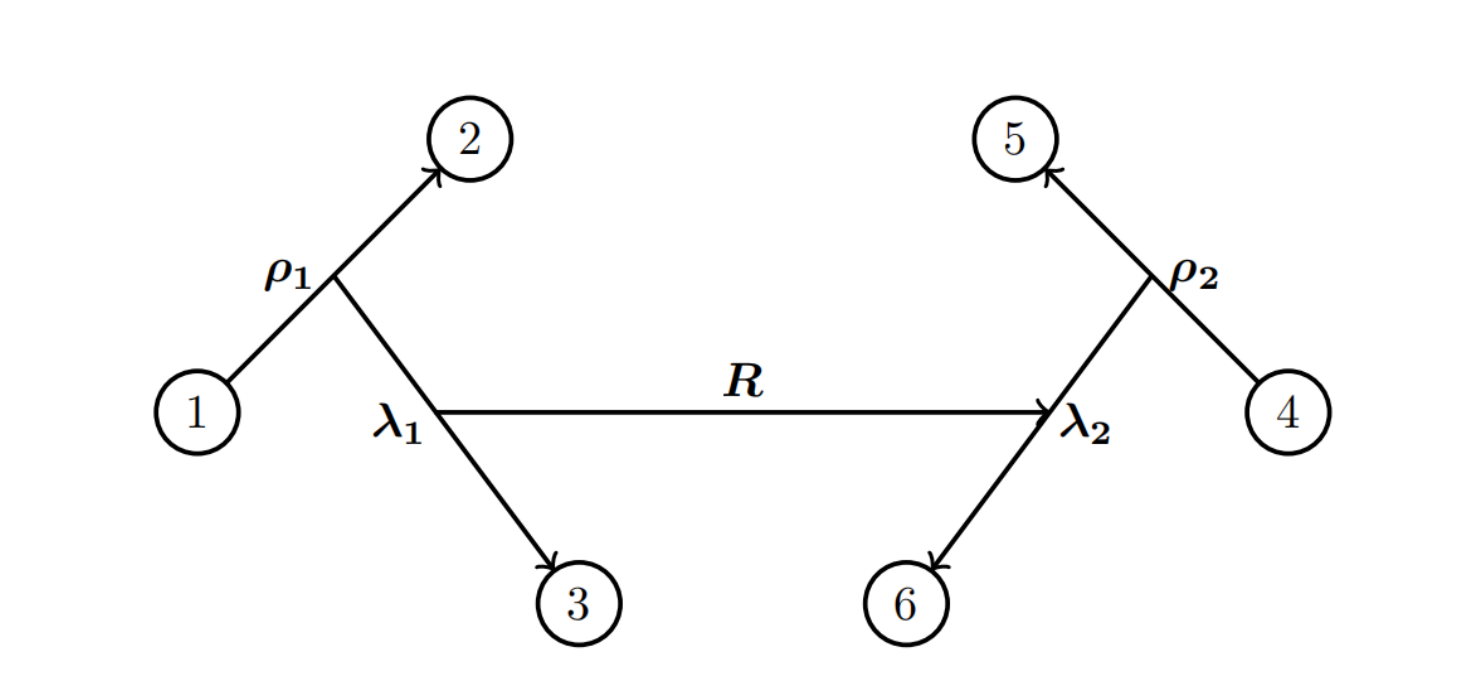} \vspace{-0.1in}
\caption{\label{fig:fig I}Jacobi coordinates of a six-quark system.}
\end{center}
\end{figure}

There are five relative motions for a six-quark system in FIG~\ref{fig:fig I}, the five Jacobi coordinates are defined as:
\begin{eqnarray}
{\boldsymbol \rho}_1    & = & {\mbox{\boldmath $r$}}_1-{\mbox{\boldmath $r$}}_2, ~~~~
{\boldsymbol \lambda}_1    = \frac{{\mbox{\boldmath $r$}}_1+{\mbox{\boldmath $r$}}_2}{2}-{\mbox{\boldmath $r$}}_3, \nonumber \\
{\boldsymbol \rho}_2    & = & {\mbox{\boldmath $r$}}_4-{\mbox{\boldmath $r$}}_5, ~~~~
{\boldsymbol \lambda}_2   = \frac{{\mbox{\boldmath $r$}}_4+{\mbox{\boldmath $r$}}_5}{2}-{\mbox{\boldmath $r$}}_6,  \\
{\boldsymbol R}    & = & \frac{{\mbox{\boldmath $r$}}_1+{\mbox{\boldmath $r$}}_2 +{\mbox{\boldmath $r$}}_3}{3}-
                         \frac{{\mbox{\boldmath $r$}}_4+{\mbox{\boldmath $r$}}_5 +{\mbox{\boldmath $r$}}_6}{3}.  \nonumber
\end{eqnarray}
${\boldsymbol r}_i$ is the position of the $i$ th particle.
The orbital wavefunctions can be written as:
\begin{eqnarray}
\Psi_{LM_L} &=& \left[ [\phi_{n_1l_1}(\boldsymbol{\rho}_1) \varphi_{n_2l_2}(\boldsymbol{\lambda}_1)]_{l_a}
 [\phi_{n_3l_3}(\boldsymbol{\rho}_2) \varphi_{n_4l_4}(\boldsymbol{\lambda}_2)]_{l_b} \right. \nonumber \\
 & & \left. \psi_{n_5l_5}(\boldsymbol{R})\right]_{LM_L}
\end{eqnarray}
where $\phi_{n_1l_1}(\boldsymbol{\rho}_1)$ represents the relative motion wave function between the quarks 1 and 2,
$\varphi_{n_2l_2}(\boldsymbol{\lambda}_1)$ indicates the relative motion between the center of mass of the quarks 1 and 2 and the quark 3.
Similarly, $\phi_{n_3l_3}(\boldsymbol{\rho}_2)$ and $\varphi_{n_4l_4}(\boldsymbol{\lambda}_2)$ stand for quarks 4,5 and 6.
$\psi_{n_5l_5}({\boldsymbol R})$ denotes the relative motion between two sub-clusters $a$ nd $b$. ``[~]" stands for the coupling of orbital angular momentum.

In the present work,
the orbital wavefunctions are fixed by solving the Schr\"{o}dinger equation with the help of GEM.
In this approach, the radial part of the orbital wave functions is expanded by a set of gaussians~\cite{GEM},
the powerful method in few body study after decades of development,
it can accurately solve the Schr\"{o}dinger equations for bound, resonant and scattering states of few-body systems~\cite{Hiyama:2018ivm}.
\begin{eqnarray}
\phi_{lm}(\mathbf{r}) & = & \sum^{n_{max}}_{n=1} c_{nl} \phi^{G}_{nlm}(\mathbf{r}) \\
\phi^{G}_{nlm}(\mathbf{r}) & = & {N}_{nl}r^{l}e^{-\nu_{n}r^{2}}\emph{Y}_{lm}(\hat{\mathbf{r}}) \\
{N}_{nl} & = & \left(\frac{2^{l+2}(2\nu_{n})^{l+3/2}}{\sqrt\pi(2l+1)!!}\right)^{\frac{1}{2}}
\end{eqnarray}
where $c_{nl}$ is the Rayleigh-Ritz variational parameter, which is determined by the dynamics of the system,
$N_{nl}$ is the normalization constant.
The Gaussian size parameters are chosen according to the following geometric progression:
\begin{eqnarray*}
\nu_{n}=\frac{1}{r^{2}_{n}}, ~~r_{n}=r_{min}a^{n-1}, ~~a=\left(\frac{r_{max}}{r_{min}}\right)^{\frac{1}{n_{max}-1}}
\end{eqnarray*}
where $n_{max}$ is the number of gaussian functions, which is determined by requiring stability of the results.

{\noindent 2. Flavor wave functions}

For sub-cluster of three $u,d$ quarks, the isospin can take 1/2 and 3/2.
Based on the flavor $SU(2)$ symmetry,
the corresponding flavor wavefunctions are:
\begin{eqnarray}
&&|\chi_{\frac12,\frac12}^{f1}\rangle  =  \sqrt{\frac{1}{6}}(2uud-udu-duu) \nonumber \\
&&|\chi_{\frac12,\frac12}^{f2}\rangle  =  \sqrt{\frac{1}{2}}(udu-duu) \nonumber \\
&&|\chi_{\frac12,-\frac12}^{f1}\rangle  =  \sqrt{\frac{1}{6}}(udd+dud-2ddu) \nonumber \\
&&|\chi_{\frac12,-\frac12}^{f2}\rangle  =  \sqrt{\frac{1}{2}}(udd-dud)  \\
&&|\chi_{\frac32,\frac32}^f\rangle  =   uuu \nonumber \\
&&|\chi_{\frac32,\frac12}^f\rangle  =  \sqrt{\frac13}(uud+udu+duu) \nonumber \\
&&|\chi_{\frac32,-\frac12}^f\rangle  =  \sqrt{\frac13}(udd+dud+ddu) \nonumber \\
&&|\chi_{\frac32,-\frac32}^f\rangle  =   ddd \nonumber
\end{eqnarray}

After consulting the Clebsch-Gordan coefficients table,
then the flavor wavefunctions for six-quark system with isospin $I=0$ are obtained
by coupling the flavor wave function of three quark system:
\begin{eqnarray*}
|\chi^{f1}_{0,0} \rangle
& = & \sqrt{\frac14}|\chi_{\frac32, \frac32}^f\rangle |\chi_{\frac32,-\frac32}^f\rangle-
      \sqrt{\frac14}|\chi_{\frac32,-\frac32}^f\rangle |\chi_{\frac32, \frac32}^f\rangle \nonumber \\
&  & -\sqrt{\frac14}|\chi_{\frac32, \frac12}^f\rangle |\chi_{\frac32,-\frac12}^f\rangle +
      \sqrt{\frac14}|\chi_{\frac32,-\frac12}^f\rangle |\chi_{\frac32, \frac12}^f\rangle \nonumber \\
|\chi^{f2}_{0,0} \rangle
& = & \sqrt{\frac12}|\chi_{\frac12,\frac12}^{f1}\rangle|\chi_{\frac12,-\frac12}^{f1}\rangle
    - \sqrt{\frac12}|\chi_{\frac12,-\frac12}^{f1}\rangle|\chi_{\frac12,\frac12}^{f1}\rangle\nonumber\\
|\chi^{f3}_{0,0} \rangle
& = & \sqrt{\frac12}|\chi_{\frac12,\frac12}^{f1}\rangle|\chi_{\frac12,-\frac12}^{f2}\rangle
     -\sqrt{\frac12}|\chi_{\frac12,-\frac12}^{f1}\rangle |\chi_{\frac12,\frac12}^{f2}\rangle\nonumber \\
|\chi^{f4}_{0,0} \rangle
& = & \sqrt{\frac{1}{2}}|\chi_{\frac12,\frac12}^{f2}\rangle|\chi_{\frac12,-\frac12}^{f1}\rangle
     -\sqrt{\frac{1}{2}}|\chi_{\frac12,-\frac12}^{f2}\rangle|\chi_{\frac12,\frac12}^{f1}\rangle\nonumber \\
|\chi^{f5}_{0,0} \rangle
& = & \sqrt{\frac{1}{2}}|\chi_{\frac12,\frac12}^{f2}\rangle|\chi_{\frac12,-\frac12}^{f2}\rangle-
      \sqrt{\frac{1}{2}}|\chi_{\frac12,-\frac12}^{f2}\rangle |\chi_{\frac12,\frac12}^{f2}\rangle \nonumber \\
\end{eqnarray*}

{\noindent 3. Spin wave function}

Due to the unique spin quantum number $S=3$ of the hexaquark system we are studying,
the spin wavefunction can be simply written as:
\begin{eqnarray}
|\chi^{\sigma 1}_{3,3} \rangle & = & \alpha\alpha\alpha\alpha\alpha\alpha .
\end{eqnarray}

{\noindent 4. Color wave function}

To construct the color wavefunctions for colorless six-quark system, there are two possible color symmetries for three-quark sub-cluster,
color singlet and color octet. All the possible wavefunctions for sub-cluster are given below.
\begin{eqnarray}
|\chi^{c_{1,1}}\rangle &=&
 ={\sqrt{\frac16}}(rgb-rbg+gbr-grb+brg-bgr) \nonumber \\
|\chi^{c_{2,1}} \rangle &=& {\sqrt{\frac16}}(2rrg-rgr-grr) \nonumber \\
|\chi^{c_{2,2}} \rangle &=& {\sqrt{\frac12}}(rgr-grr) \nonumber \\
|\chi^{c_{3,1}} \rangle &=& {\sqrt{\frac16}}(2rrb-rbr-brr) \nonumber \\
|\chi^{c_{3,2}} \rangle &=& {\sqrt{\frac12}}(rbr-brr) \nonumber\\
|\chi^{c_{4,1}} \rangle &=& {\sqrt{\frac16}}(rgg+grg-2ggr) \nonumber \\
|\chi^{c_{4,2}} \rangle &=& {\sqrt{\frac12}}(rgg-grg)\nonumber \\
|\chi^{c_{5,1}} \rangle &=& {\sqrt{\frac{1}{12}}}(2rgb-rbg+2grb-gbr-brg-bgr) \nonumber \\
|\chi^{c_{5,2}} \rangle &=& {\sqrt{\frac14}}(rbg-gbr+brg-bgr)\nonumber \\
|\chi^{c_{6,1}} \rangle &=& {\sqrt{\frac{1}{12}}}(2rgb+rbg-2grb-gbr-brg+bgr) \nonumber \\
|\chi^{c_{6,2}} \rangle &=& {\sqrt{\frac14}}(rbg+gbr-brg-bgr) \nonumber\\
|\chi^{c_{7,1}} \rangle &=& {\sqrt{\frac16}}(rbb+brb-2bbr) \nonumber \\
|\chi^{c_{7,2}} \rangle &=& {\sqrt{\frac12}}(rbb-brb) \nonumber \\
|\chi^{c_{8,1}} \rangle &=& {\sqrt{\frac16}}(2ggb-gbg-bgg)\nonumber \\
|\chi^{c_{8,2}} \rangle &=& {\sqrt{\frac12}}(gbg-bgg)\nonumber \\
|\chi^{c_{9,1}} \rangle &=& {\sqrt{\frac16}}(gbb+bgb-2bbg) \nonumber \\
|\chi^{c_{9,2}} \rangle &=& {\sqrt{\frac12}}(gbb-bgb) \nonumber
\end{eqnarray}
By using CG coefficients of $SU(3)$, then we get color singlet-singlet and color octet-octet wavefunctions of a six-quark system:
\begin{eqnarray}
|\chi^{c}_1 \rangle &=& |\chi^{c_{1,1}} \rangle |\chi^{c_{1,1}} \rangle\nonumber \\
|\chi^{c}_2 \rangle &=& {\sqrt{\frac18}}(|\chi^{c_{2,1}} \rangle |\chi^{c_{9,1}} \rangle-
                                         |\chi^{c_{3,1}} \rangle |\chi^{c_{8,1}} \rangle \nonumber \\
                                      &-&|\chi^{c_{4,1}} \rangle |\chi^{c_{7,1}} \rangle+
                                         |\chi^{c_{5,1}} \rangle |\chi^{c_{5,1}} \rangle-
                                         |\chi^{c_{8,1}} \rangle |\chi^{c_{3,1}} \rangle\nonumber \\
                                     &+& |\chi^{c_{6,1}} \rangle |\chi^{c_{6,1}} \rangle-
                                         |\chi^{c_{7,1}} \rangle |\chi^{c_{4,1}} \rangle
                                        +|\chi^{c_{9,1}} \rangle |\chi^{c_{2,1}} \rangle)\nonumber \\
|\chi^{c}_3 \rangle &=& {\sqrt{\frac18}}(|\chi^{c_{2,1}} \rangle |\chi^{c_{9,2}} \rangle-
                                         |\chi^{c_{3,1}} \rangle |\chi^{c_{8,2}} \rangle \nonumber \\
                                     &-& |\chi^{c_{4,1}} \rangle |\chi^{c_{7,2}} \rangle
                                        +|\chi^{c_{5,1}} \rangle |\chi^{c_{5,2}} \rangle-
                                         |\chi^{c_{8,1}} \rangle |\chi^{c_{3,2}} \rangle \nonumber \\
                                     &+& |\chi^{c_{6,1}} \rangle |\chi^{c_{6,2}} \rangle
                                       - |\chi^{c_{7,1}} \rangle |\chi^{c_{4,2}} \rangle
                                        +|\chi^{c_{9,1}} \rangle |\chi^{c_{2,2}} \rangle)\nonumber \\
|\chi^{c}_4 \rangle &=& {\sqrt{\frac18}}(|\chi^{c_{2,2}} \rangle |\chi^{c_{9,1}} \rangle-
                                         |\chi^{c_{3,2}} \rangle |\chi^{c_{8,1}} \rangle \nonumber \\
                                     &-& |\chi^{c_{4,2}} \rangle |\chi^{c_{7,1}} \rangle
                                        +|\chi^{c_{5,2}} \rangle |\chi^{c_{5,1}} \rangle-
                                         |\chi^{c_{8,2}} \rangle |\chi^{c_{3,1}} \rangle\nonumber \\
                                     &+& |\chi^{c_{6,2}} \rangle |\chi^{c_{6,1}} \rangle-
                                         |\chi^{c_{7,2}} \rangle |\chi^{c_{4,1}} \rangle
                                        +|\chi^{c_{9,2}} \rangle |\chi^{c_{2,1}} \rangle)\nonumber \\
|\chi^{c}_5 \rangle &=& {\sqrt{\frac18}}(|\chi^{c_{2,2}} \rangle |\chi^{c_{9,2}} \rangle
                                        -|\chi^{c_{3,2}} \rangle |\chi^{c_{8,2}} \rangle\nonumber \\
                                     &-& |\chi^{c_{4,2}} \rangle |\chi^{c_{7,2}} \rangle
                                        +|\chi^{c_{5,2}} \rangle |\chi^{c_{5,2}} \rangle
                                        -|\chi^{c_{8,2}} \rangle |\chi^{c_{3,2}} \rangle\nonumber \\
                                     &+& |\chi^{c_{6,2}} \rangle |\chi^{c_{6,2}} \rangle-
                                         |\chi^{c_{7,2}} \rangle |\chi^{c_{4,2}} \rangle
                                        +|\chi^{c_{9,2}} \rangle |\chi^{c_{2,2}} \rangle)\nonumber
\end{eqnarray}

To save space, the detailed color wavefunctions are omitted here.

Finally, the total wave function of the six-quark system is written as:
\begin{eqnarray}
\Psi_{JM_J}^{i,j,k} &=& {\cal A} \left[ \left[
 \psi_{L}\chi^{\sigma_i}_{S}\right]_{JM_J}
   \chi^{f}_j \chi^{c}_k \right], \\
&& (i=1\sim 1,~j=1\sim 5,~k=1\sim 5), \nonumber
\end{eqnarray}
where $J$ is the total angular momentum and $M_{J}$ is the 3rd component of the total angular momentum,
$\cal{A}$ is the antisymmetry operator of the system, it consists of three parts,
${\cal A}_{123}$,~${\cal A}_{456}$ represent the antisymmetry operator of the sub-cluster $a$ and $b$, respectively,
${\cal A}_{123,456}$ stand for the antisymmetry operator between the two sub-clusters,
${\cal A}_{123,456}$ is obtained by operating a coset decomposition of $S_6\supset S_3\otimes S_3$ to find the coset representative.
\begin{equation*}
\begin{aligned}
 &{\cal A}_{123}=\sqrt{\frac{1}{6}}[1-(13)-(23)][1-(12)]\\
 &{\cal A}_{456}=\sqrt{\frac{1}{6}}[1-(46)-(56)][1-(45)]
\end{aligned}
\end{equation*}
\begin{equation*}
\begin{aligned}
{\cal A}_{123,456}=\sqrt{\frac{1}{20}}[1&-(14)-(15)-(16)\\
                                        &-(24)-(25)-(26)\\
                                        &-(34)-(35)-(36)\\
                                    &+(14)(25)+(14)(26)\\
                                    &+(14)(35)+(14)(36)\\
                                    &+(15)(26)+(15)(36)\\
                                    &+(24)(35)+(24)(36)\\
                                    &+(25)(36)-(14)(25)(36)]
\end{aligned}
\end{equation*}
\begin{equation*}
 {\cal A}= {\cal A}_{123,456}{\cal A}_{456}{\cal A}_{123}
\end{equation*}

The eigen-energy of system is obtained by solving the following eigen-equation:
\begin{equation*}
H\Psi_{JM_J}=E\Psi_{JM_J},
\end{equation*}
by using Rayleigh-Ritz variational principle.

\begin{table}
\caption{\label{tab:table2} The possible channels of the hexaquark system with $IJ^P=03^+$. The subscripts ``1" and ``8" means color
singlet and color octet, respectively. The superscripts ``S'' and ``A'' denote the permutation symmetry of first two quarks in each sub-cluster of the flavor
wave functions. The superscripts ``4" is $2S+1$, $S$ is the spin of the sub-cluster.}
\begin{tabular}{ccccccc}
\toprule
Index &${c_i \sigma_j f_k}$ & Physical content &  \\ \hline
1& $i=1;j=1;k=1$ & $^{4}\Delta_1{~}^{4}\Delta_1$     \\
2& $i=1;j=1;k=2$ & ${^{4}N^S_1}{~}^{4}N^S_1$       \\
3& $i=2;j=1;k=5$ & ${^{4}N^A_8}{~}^{4}N^A_8$       \\
4& $i=3;j=1;k=4$ & ${^{4}N^A_8}{~}^{4}N^S_8$       \\
5& $i=4;j=1;k=3$ & ${^{4}N^S_8}{~}^{4}N^A_8$       \\
6& $i=5;j=1;k=1$ & $^{4}\Delta_8{~}^{4}\Delta_8$     \\
7& $i=5;j=1;k=2$ & ${^{4}N^S_8}{~}^{4}N^S_8$       \\
\botrule
\end{tabular}
\end{table}

In the present work,
we investigate the hexaquark systems with quantum numbers $IJ^P=03^+$ in the quark model.
We are interested in the low energy states of the hexaquark systems,
so here we set all the orbital angular momenta to be zero.
All possible configurations for flavor, spin,
and color degrees of freedom are considered.
The possible channels of the two configurations are listed in Table~\ref{tab:table2}.
The first channel is the color-singlet-singlet one, others are hidden-color ones.

\begin{table}[h]
\caption{\label{tab:table3} Quark model parameters}
\begin{tabular}{c|cc cc}
\toprule
                        &                            & NQM     & CHQM    \\  \hline
Quark                  &$m_u$ (MeV)                  &313      &~~313    \\
masses                 &$m_d$ (MeV)                  &313      &~~313    \\  \hline
                       &$\Lambda_\pi$ (fm$^{-1}$)    &-        &~~4.20   \\
                       &$\Lambda_\sigma$ (fm$^{-1}$) &-        &~~4.20   \\
                       &$m_\pi$ (fm$^{-1}$)          &-        &~~0.70   \\
Goldstone              &$m_\sigma$ (fm$^{-1}$)       &-        &~~3.42   \\
bosons                 &$g^2_{ch}/(4\pi)$            &-        &~~0.54   \\
                       &$\theta_P(^\circ)$           &-        &~~-15    \\  \hline
                       &$a_c$ (MeV$\cdot$ fm$^{-2}$) &36.94    &~~36.94  \\
Confinement            &$V_0$ (MeV)                  &30.93    &~~10.2   \\
                       &$\alpha_{uu}$                &0.66     &~~0.68   \\  \hline
  OGE                  &$\hat{r}_0~$(MeV~fm)         &16.8     &~~13.7   \\
\botrule
\end{tabular}
\end{table}
\section{Results and discussions}
The model parameters fixed by fitting baryon spectra are listed in Table~\ref{tab:table3}.
Our main purpose is to calculate the energy of non-strange dibaryon $d^*$,
so we only list light baryon spectra here. The baryon masses are obtained by solving the three-body Schr\"{o}dinger
equation by using GEM.
From the Tables~\ref{tab:table4} and~\ref{tab:table5},
one can see that the results are stable with gaussian number $n=7$.

\begin{table}[h]
\caption{\label{tab:table4} light baryon under na\'ive quark model}
\begin{tabular}{ccccccc}
\toprule
n                               & 4        & 5          & 6      & 7      & 8      & 9       \\ \hline
$N$ (MeV)                       & 975.1    & 938.3      & 936.4  & 936.3  & 935.9  & 935.6   \\
$\Delta$ (MeV)                  & 1310.4   & 1242.9     & 1232.8 & 1232.3 & 1232.3 & 1232.3  \\
\botrule
\end{tabular}
\end{table}
\begin{table}[h]
\caption{\label{tab:table5} baryon under chiral quark model}
\begin{tabular}{ccccccc}
\toprule
n                               & 4        & 5          & 6      & 7      & 8      & 9       \\ \hline
$N$ (MeV)                       & 968.0    & 939.3      & 938.0  & 937.6  & 936.6  & 935.5   \\
$\Delta$ (MeV)                  & 1283.9   & 1232.7     & 1224.6 & 1223.8 & 1223.8 & 1223.7  \\
\botrule
\end{tabular}
\end{table}

For six-quark system,
five relative motions needed to be expanded by a set of Gaussians,
so the dimension of matrix to be diagonized is very large.
For single channel calculation,
the dimension of the matrix is $7^5=16807$ for $n_{max}$=7,
the dimension of the full channel coupling calculation will be $7^6=117649$,
it is beyond our ability.
So for the full-channel coupling calculation, we set $n_{max}$=6.

To check the contributions of hidden-color channels,
we first do a single channel calculation,
only consider the color singlet-singlet channel.
Then do a full channel coupling calculation,
and comparing two results,
one can see the contribution of the hidden-color channels.
The calculated results of $IJ^P=03^+$ are given in Table~\ref{tab:table6} and Table~\ref{tab:table7}.
$E^{\rm Theo}_{th}$ means the theoretical thresholds (the sum of the masses of two $\Delta$),
$E_{sc}$, $E_{cc}$ represent the the lowest energies for single channel and full channel coupling calculations, respectively.
$B_{sc}$, $B_{cc}$ represent the corresponding binding energies.

\begin{table}[h]
\caption{\label{tab:table6} The energy of the hexaquark system with $IJ^P=03^+$ in na\'ive quark model. ``sc" and ``cc" means single channel and
full channel coupling.}
\begin{tabular}{ccccccc}
\toprule
$n_{max}$                                & 4       & 5        & 6      & 7       \\ \hline
$E^{\rm Theo}_{th}$ (MeV)                & 2620.8  & 2485.8   & 2465.7 & 2464.5  \\
$E_{sc}$ (MeV)                           & 2606.2  & 2474.4   & 2454.1 & 2453.3  \\
$B_{sc}$ (MeV)                           & 14.6    & 11.4     & 11.6   & 11.2    \\
$E_{cc}$ (MeV)                           & 2582.9  & 2469.1   & 2450.9           \\
$B_{cc}$ (MeV)                           & 37.9    & 16.7     & 14.8             \\
\botrule
\end{tabular}
\end{table}

\begin{table}[h]
\caption{\label{tab:table7} The energy of the hexaquark system with $IJ^P=03^+$ in chiral quark model. ``sc" and ``cc" means single channel and
full channel coupling.}
\begin{tabular}{ccccccc}
\toprule
n                                     & 4        & 5          & 6      & 7        \\ \hline
$E^{\rm Theo}_{th}$ (MeV)             & 2567.8   & 2465.4     & 2449.2 & 2447.6   \\
$E_{sc}$ (MeV)                        & 2470.0   & 2379.0     & 2368.0 & 2367.0   \\
$B_{sc}$ (MeV)                        & 97.8     & 86.4       & 81.2   & 80.6     \\
$E_{cc}$ (MeV)                        & 2432.4   & 2361.3     & 2353.4            \\
$B_{cc}$ (MeV)                        & 135.4    & 104.1      & 95.8              \\
\botrule
\end{tabular}
\end{table}

\begin{table}[h]
\caption{\label{radius} The RMS radius of the hexaquark system with $IJ^P=03^+$ in na\'ive quark model.}
\begin{tabular}{c|cccccc}
\toprule
            &        & n                     & 4       & 5        & 6      & 7       \\  \hline
            &        & $ r(d^*)   $ (fm)     & 1.30    & 1.20     & 1.10   & 1.10    \\
  na\'ive   &   sc   & $ r(\Delta)$ (fm)     & 0.51    & 0.51     & 0.51   & 0.51    \\
quark model &        & $ r(N)     $ (fm)     & 0.67    & 0.64     & 0.64   & 0.64    \\  \cmidrule{2-7}
            &   cc   & $ r(d^*)   $ (fm)     & 1.20    & 1.10     & 1.10             \\  \hline

            &        & $ r(d^*)   $ (fm)     & 0.87    & 0.85     & 0.86   & 0.86    \\
 chiral     &   sc   & $ r(\Delta)$ (fm)     & 0.69    & 0.65     & 0.65   & 0.65    \\
quark model &        & $ r(N)     $ (fm)     & 0.50    & 0.50     & 0.50   & 0.50    \\  \cmidrule{2-7}
            &   cc   & $ r(d^*)   $ (fm)     & 0.86    & 0.83     & 0.85             \\

\botrule
\end{tabular}
\end{table}

In na\'ive quark model,
the single channel calculation shows that the binding energy approaches to 11.2 MeV,
the full channel coupling lower the energy of the system a little.
From the energy,
one can see the contributions of hidden-color channels to the energy of the system is small,
less than 10\%.
Huang's updated preliminary results
show that the binding energy of $\Delta\Delta$ system is $18$ MeV,
when the channel coupling of $\Delta\Delta$ and $CC$ is further considered,
the binding energy of the system is found to be 21 MeV~\cite{Huang:2022qgx}.
The percentage of color singlet-singlet channel dominates the state in the full channel coupling calculation confirms the results.
Compared to the experimental data,
binding energy is about $80$ MeV,
na\'ive quark model obtains a smaller binding energy,
which infers that the attraction provided by one-gluon-exchange is not enough.

In the chiral quark model,
because of the introducing of the $\sigma$-meson exchange,
the attraction between two sub-clusters are rather strong,
which leads a large binding energy,
around 80 MeV for single channel calculation,
96 MeV for the full channel coupling calculation,
the difference is about 15 MeV,
it indicates that the hidden color channel is not so important again,
and the binding energy is similar to the experimental value.

To find the structure of the state,
the root-mean-square (RMS) radius of the system is calculated,
which are shown in Table \ref{radius}.
For comparison,
the RMS radius of nucleon and $\Delta$ are also given.
The radius of the system is defined as:
\begin{eqnarray}
\mathbf{r}(q^3) & = & \mathbf{r}_1-\frac{1}{3}\sum_{i=1}^3 \mathbf{r}_i , ~~~~{\mbox{for 3-quark system}}, \\
\mathbf{r}(q^6) & = & \mathbf{r}_1-\frac{1}{6}\sum_{i=1}^6 \mathbf{r}_i , ~~~~{\mbox{for 6-quark system}}.
\end{eqnarray}

Since they are all identical particles,
the distance between any quark and the center of mass is defined as the radius of a three body or six body system.

From the Table \ref{radius},
one can see that the radius of $d^*$ is around 1.1 fm (in na\'ive quark model) or 0.85 fm (in chiral quark model),
whereas the radius of $\Delta$ is around 0.51 fm and 0.65 fm in two quark models, respectively.
The channel coupling has tiny effect on the radius of the system.
By comparing the results of two quark models,
it is clear that the larger the binding energy,
the smaller the radius.
The radius show that the state $d^*$ may be a compact object.

\section{Summary}
We investigated the nonstrange hexaquark state with quantum numbers $IJ^P=03^+$ in the framework of quark models.
To conduct a precise calculation, GEM is employed.

Our calculation results show that the state $d^*(2380)$ is a compact object,
and the color singlet-singlet channel dominates the state.
The hidden-color component can lower the energy of the state a little,
less than 10\%.
The results are basically consistent with that of Huang's updated calculation~\cite{Huang:2022qgx},
the influence of hidden-color channels in $d^*$ may not be significant.
And our binding energy under chiral quark model approaching experimental values,
the radius of $d^*(2380)$ under two quark models is $0.8-1.1$ fm,
consistent with the outcome obtained from lattice QCD,
their typical of size of the quasi-bound state is $0.8-1$ fm
and final value of the binding energies read $25-40$ MeV below the $\Delta\Delta$ threshold~\cite{Gongyo:2020pyy}.
QCD sum rule obtained $M_{d^*}=2.4\pm0.2$ GeV,
but they thought it is challenging to determine whether the $d^*(2380)$ is a $\Delta$-$\Delta$ bound state or a six-quark state with hidden-color configurations~\cite{Chen:2014vha}.
C.S. An {\em et al} calculate the energies of the genuine hexaquark configurations,
considering instanton-induced hyperfine interaction between quarks~\cite{An:2016txh}.

By combining three diquarks of both types ($\bar{\mathbf{3}}_c, I=1$) or ($\mathbf{6}_c, I=0$),
Kim {\em et al} demonstrated that hexaquark picture is promising for $d^*(2380)$~\cite{Kim:2020rwn}.
B. Kabirimanesh {\em et al} considered that dibaryon are consisted of three diquarks,
and obtained their estimation of hexaquark mass $ M_H \simeq2332$ MeV,
about $48$ MeV lower than the experimental value $d^*(2380)$~\cite{Kabirimanesh:2022hrp}.
Other calculations based on the diquark model have achieved results close to the experimental values~\cite{Gal:2019xju,Shi:2019dpo}.
Determining whether $d^*(2380)$ is a six quark dominated state is of great significance,
it may imply a new degree of freedom and enables us to better understand baryon-baryon interactions~\cite{Kukulin:2020gee,Tursunbayev:2020dif}.

The focus of this paper is to perform dynamic calculation of $d^*(2380)$ without assuming any presupposed structure,
so we can study the size, structure, and mass of it ,
next, the magnetic moment, quadrupole and octupole deformations, dacay width needed to be discussed,
some work has been completed,
M. Bashkanov calculated the quadrupole and octupole moments in a pion cloud model~\cite{Bashkanov:2019mbz},
the result is in agreement with that obtained by resonating group method~\cite{Dong:2018emq}.
For the unusual narrow decay width of $d^*(2380)$,
a free $\Delta$'s decay width is $\Gamma_\Delta \sim 115$ MeV,
but $\Gamma_{d^*} \sim 70$ MeV,
distortion of $\Delta$ wavefunction and the phase space constrain can give the answer,
which is our next work.

A lesson from the calculation is that for the compact object,
to freeze the internal structure of sub-cluster to simplify the calculation is not a good idea,
especially for precise calculation.
In the present paper,
we discuss the most fascinating non-strange dibaryons in the light quark sector.
The dibaryons with $s$ quark, $c$ quark and $b$ quark are also of interest to us,
because these states are accessible in experiments,
we may consider these hexaquark states in the future work.
It is of great significance to determine whether there are stable dibaryon systems except deuteron.
With the development of experiment and theory,
we should have confidence in the future of hexaquark,
hence, more efforts are needed~\cite{Bashkanov:2018bqs}.

\section*{Acknowledgments}
This work is supported partly by the National Natural Science Foundation of China under
Contract Nos. 11775118, and 11535005.

\end{document}